\documentclass{article}

\usepackage{microtype}
\usepackage{graphicx}
\usepackage{subcaption}
\usepackage{booktabs} 

\usepackage{hyperref}

\usepackage[preprint]{icml2026}

\usepackage{amsmath}
\usepackage{amssymb}
\usepackage{mathtools}
\usepackage{amsthm}

\newcommand\para[1]{\noindent\textbf{#1}}

\usepackage[capitalize,noabbrev]{cleveref}

\theoremstyle{plain}

\theoremstyle{definition}

\theoremstyle{remark}

\usepackage[textsize=tiny]{todonotes}

\usepackage{enumitem}

\icmltitlerunning{Amalgam: Hybrid LLM-PGM Synthesis Algorithm for Accuracy and Realism}

\begin{document}

\twocolumn[
    \icmltitle{Amalgam: Hybrid LLM-PGM Synthesis Algorithm for Accuracy and Realism}



    \icmlsetsymbol{equal}{*}

    \begin{icmlauthorlist}
        \icmlauthor{Antheas Kapenekakis}{aau}
        \icmlauthor{Bent Thomsen}{aau}
        \icmlauthor{Katja Hose}{tuw}
        \icmlauthor{Michele Albano}{aau}
    \end{icmlauthorlist}

    \icmlaffiliation{aau}{Aalborg University, Aalborg, Denmark}
    \icmlaffiliation{tuw}{TU Wien, Vienna, Austria}

    \icmlcorrespondingauthor{Antheas Kapenekakis}{antheas@cs.aau.dk}

    \icmlkeywords{Machine Learning, Synthetic Data, LLM}

    \vskip 0.3in
]



\printAffiliationsAndNotice{}  

\begin{abstract}
    To generate synthetic datasets, e.g., in domains such as healthcare, the literature proposes approaches of two main types: Probabilistic Graphical Models (PGMs) and Deep Learning models, such as LLMs.
    While PGMs produce synthetic data that can be used for advanced analytics, they do not support complex schemas and datasets.
    LLMs on the other hand, support complex schemas but produce skewed dataset distributions, which are less useful for advanced analytics.
    In this paper, we therefore present Amalgam, a hybrid LLM-PGM data synthesis algorithm supporting both advanced analytics, realism, and tangible privacy properties.
    We show that Amalgam synthesizes data with an average 91 \% $\chi^2 P$ value and scores 3.8/5 for realism using our proposed metric, where state-of-the-art is 3.3 and real data is 4.7. 
\end{abstract}

\section{Introduction}
Data synthesis is the process of generating artificial data that mimics the
properties of a real dataset.
It is widely used in multiple domains, with two main purposes:
augmenting training data for machine learning and sharing data in a
privacy-preserving manner.
In this paper, we focus on the latter, which is especially relevant in
domains that handle confidential data, such as healthcare and finance.
In these domains, algorithms focus on two approaches:
Probabilistic Graphical Models (PGMs) and Deep Learning Networks.

Probabilistic Graphical Models (PGMs)~\cite{privlava,
    caiDataSynthesisDifferentially2021,mckennaAIMAdaptiveIterative2022}
are low parameter statistical models that learn the joint distribution of
a dataset.
Because of their structure, they are interpretable and the resulting data
has strong analytical properties (e.g., Kullback-Leiber divergence is
minimal).
In addition, due to their low parameter count, they have excellent
privacy properties, especially when combined with methods such as
Differential Privacy (DP)~\cite{dwork2006differential}, and through
work such as the system Pasteur~\cite{pasteur}, they are shown to scale
to very large datasets.
However, PGMs do not scale to complex datasets, with most work focusing
on tabular data~\cite{caiDataSynthesisDifferentially2021,
    mckennaAIMAdaptiveIterative2022,zhangPrivBayesPrivateData2017b,
    mckennaWinningNISTContest2021d}.

On the other hand, Deep Learning models~\cite{pategan,medgan}, especially
Large Language Models (LLMs)~\cite{llmframework,llmbalance,llmenhance},
due to their high parameter count and being pre-trained on a large corpus
of data, have the potential to handle complex datasets, which may contain
sequences or complex relational structures.
However, due to their high parameter count, if the input dataset is used
for training, it is hard to ensure that the model does not expose the
training dataset and that it is privacy-preserving without sacrificing
accuracy~\cite{dpsgd}.
Moreover, the resulting data tends to capture a less accurate
distribution~\cite{taoBenchmarkingDifferentiallyPrivate2021}

These approaches appear complimentary: PGMs provide accurate analytics and
privacy, while LLMs provide the ability to handle complex datasets.
However, combining them and retaining these qualities remains an unanswered
question.
Evaluating the result for realism and analytics accuracy without a human in the loop remains
an open question as well.
In this paper, we aim to answer the following research questions:
\begin{enumerate}[start=1,label={(\bfseries RQ\arabic*):},leftmargin=20pt]
    \item How do we combine a PGM and an LLM to produce synthetic data where the following are met:
          \begin{enumerate}[start=1,label={(\bfseries Sub-RQ\arabic*):},leftmargin=30pt]
              \item We preserve the accuracy of analytics from the PGM? (accuracy)
              \item The foundational knowledge of an LLM is used to handle and
                    enhance complex dataset? (complexity)
              \item The resulting data features quantified privacy properties? (privacy)
              \item The computational complexity remains reasonable to perform locally? (efficiency)
          \end{enumerate}
    \item Given a relational dataset as a result, how do we evaluate the synthetic data for:
          \begin{enumerate}[start=1,label={(\bfseries Sub-RQ\arabic*):},leftmargin=30pt]
              \item Realism, i.e., how lifelike is the data?, in an unattended fashion,
                    i.e., without review by a human (realism)
              \item Accuracy of analytics, i.e., how well key statistics
                    are preserved?, across tables in a standardized manner (accuracy)
          \end{enumerate}
\end{enumerate}

In this paper, we aim to combine the benefits of PGMs and LLMs, through
introducing Amalgam, a hybrid LLM-PGM synthesis algorithm.
Amalgam uses a PGM to learn the joint distribution of the dataset and generate
summary statistics which are privacy-preserving and accurate.
The summary statistics are used to condition an LLM, along similar samples from
the original dataset, to generate synthetic samples in a zero-shot manner.
The resulting data features a favorable combination of the qualities of
privacy-preservation, accuracy, complexity of input data, and computational
efficiency over state-of-the-art.
Then, we present an unattended evaluation method for dataset realism, where an
LLM is used to reason and then score how lifelike each sample is.
Finally, we present experiments where we compare Amalgam against state-of-the-art
on two public datasets for accuracy and realism.
We show that Amalgam presents a favorable combination of qualities over
state-of-the-art (i.e., higher privacy and accuracy than LLMs, handling higher complexity
with higher realism than PGMs).

\section{Related Work}
Work on data synthesis spans multiple domains and data types, where foundation of
this paper is relational data synthesis and privacy-aware synthesis.
For structured data, most work focuses on tabular datasets and on
two approaches.
The first one is PGM models~\cite{mckennaAIMAdaptiveIterative2022,caiDataSynthesisDifferentially2021,
    zhangPrivBayesPrivateData2017b,mckennaWinningNISTContest2021d,privpgm},
which feature the Differential Privacy technique~\cite{dwork2006differential}
to ensure privacy.
The second one is Deep Learning models~\cite{pategan,medgan,xuModelingTabularData2019}.
Most of these works involve the use of a Generative Adversarial Network
(GAN) or a Variational Autoencoder (VAE).
In benchmarks~\cite{taoBenchmarkingDifferentiallyPrivate2021}, PGM models
tend to outperform Deep Learning models in terms of distributional accuracy
on tabular datasets.
Moreover, training Deep Learning models via Differential Privacy tends to
result in a significant accuracy loss as shown in those
papers~\cite{pategan,medgan,xuModelingTabularData2019} and since tabular
datasets are trivial, the ability of Deep Learning models to handle
complex datasets is not leveraged.

Recently, two PGM works have attempted to synthesize relational
data~\cite{privlava,mare} using PGMs.
PrivLava~\cite{privlava} derives from PrivMRF~\cite{caiDataSynthesisDifferentially2021}
to allow generating multiple disjoint tables together, without inter-table
dependencies (e.g., two subsequent patient admissions in a medical dataset
are generated independently).
MARE~\cite{mare} is PGM algorithm agnostic and models a relational dataset
as a template representation graph~\cite{pgm-koller}.
This graph can model three types of child-to-parent table relationships:
sequential (via Markov Chains), associative (via a technique called
unrolling), and independent, where generated rows can reference all
previously generated values (e.g., an admission column can reference
the patient's sex).
Both of these approaches are limited to simple relational datasets and
scaling to more complex dataset requires significant engineering.
This is similar to early attempts at computer vision, which used complex
rulesets to e.g., recognize faces before being obsoleted by Deep Learning.

LLM models are used in general for generation purposes~\cite{transformer},
and for synthetic data~\cite{llmframework}.
Recently, great strides have been made on models that can run on-premises,
in secure infrastructure~\cite{llmquant,deepseek-v3,gemma-3}.
These models can be effectively run on a single GPU with an adequate token
generation speed (hundreds of tokens per second).
However, small models are regarded as having a lower quality output than
higher end proprietary models.

To aid in augmenting data quality, recent works suggest a grammar based
approach~\cite{llm-grammar-syncode, llm-grammar-grammarllm}to limit the output
tokens of the models to a specific schema (e.g., SQL, JSON, XML).
Both local LLM inference frameworks (e.g., llamma\_cpp, vllm) and online
providers (e.g., OpenAI) provide an approach to limit the output
tokens with a specific grammar.


\section{Amalgam Algorithm}
In this section, we present the algorithm Amalgam.
First, we present the two main steps of the algorithm: structure learning
(i.e., training the PGM and preparing the LLM), and sampling (i.e., generating
synthetic data).
Then, we present an evaluation method for determining the realism of synthetic data
based on an assessment by an LLM agent.

\subsection{Structure Learning}\label{sec:fitting}
Structure learning begins by collecting metadata per entity in the dataset
to form a tabular dataset distilling their analytics.
An entity is defined as a single row in the main table of the dataset.
For example, in a medical dataset, an entity may be a patient or an
admission in the intensive care unit (ICU) if the dataset does not link to patients.
In such cases, the main table is the patient table or the admission table, respectively.
The analytics are defined as key statistics that we want to preserve in
the synthetic data (e.g., the age distribution, treatment length, or medicine
distribution).

To create the tabular dataset of analytics, first, the relational dataset
is discretized to ensure all columns are categorical.
Numerical columns are binned into a set of discrete ranges.
Integers are used, selecting the smallest unsigned integer width that fits
the domain of the column (e.g., uint8 for values between 0 and 255).
The core of the analytics table is formed by the main table of the dataset.
Then, through a series of left joins, other tables in the dataset are
merged into the main table, selecting one row per entity.
If the relationship between the tables is sequential (e.g., admissions
sorted by date), the first row is selected.
Otherwise, an arbitrary row is chosen.
Moreover, a count column is added per child table to indicate how many
rows the original entity had.

The analytics tabular dataset is used to train a PGM model,
e.g., PrivBayes~\cite{zhangPrivBayesPrivateData2017b} or
PrivMRF~\cite{caiDataSynthesisDifferentially2021}, which learns the joint
distribution of the analytics.
To ensure privacy when generating analytics, privacy-aware algorithms utilize
the mechanism Differential Privacy~\cite{dwork2006differential}.
By using a low privacy budget parameter $\epsilon$, e.g., 2, the algorithms
ensure the analytics, which reference the whole dataset, are protected from
inference.
The resulting PGM is used to generate samples of the analytics, which
will be used to condition the LLM during sampling.
By avoiding the use of fine tuning or LLMs during structure learning,
we avoid the privacy and performance implications these processes
carry.

\subsection{Sampling}\label{sec:sampling}
During sampling, the PGM generates conditioning values (e.g., age, sex).
These values are used with a similarity function to collect a number
of top samples in the original data to be provided as a reference for
synthesis.
Then, the analytics are converted into a list of human readable values
(e.g., ``age: 45'', ``entry time: 10:30'') and provided along-side the similar
samples in a JSON format.
These two components are used to form the prompt for the LLM, which
is bound by a grammar to generate a JSON object that mirrors the dataset
schema.
The output of the LLM is parsed and returned as the synthetic sample.

\para{Similarity Function.}
We define the similarity function as a function
that receives as input
the single column histograms of the analytics dataset, measured during
the fitting of the PGM, and two samples, $a$, $b$.
It can be formulated as follows:
\begin{equation}
    similar(a, b) = \sum_{i=1}^{n} (a_i = b_i) \cdot \frac{1}{p_i}
\end{equation}
Where $n$ is the number of columns, $a_i$ and $b_i$ are the values of
column $i$ in samples $a$ and $b$ respectively, and $p_i$ is the
probability of the value $a_i$ in the marginal of column $i$.
An alternate representation of a single-column marginal is a
histogram of the value counts of a column.
This heuristic representation increases the perceived similarity between
two samples proportionally to the number of values they have common
and how rare these values are.

The similarity scores are generated per conditioning sample and row in
the original data (complexity $O(MN)$, where $N$ is the number of rows in
the original and $M$ is the number of conditioning samples).
If $M = N$, this would result in a complexity of $O(N^2)$.
However, as we show in the experimental section, the computational cost
of each sampled row requires running the LLM, where the cost is
$O(MF_{\textrm{LLM}})$, where $F_{\textrm{LLM}}$ is the cost of running
the LLM per sample.
As $F_{\textrm{LLM}}$ is significantly higher than $N$ similarity function
calculations, $M << N$ must hold for synthesis to finish in a realistic
timeframe.
Therefore, the complexity of the similarity function is not a bottleneck
in the overall synthesis process.

\para{LLM Synthesis.}
First, the original dataset schema is converted into a JSON schema.
This is done by mapping each row of the dataset to a dictionary, and modelling
dependencies between tables as lists of those dictionaries.
The conversion is performed by parsing the dataset schema recursively.
Then, the $n$ samples are converted into this schema, by using human-readable
values where possible (e.g., time value ``5'' becomes ``4:00'').

The samples and the conditioning values are used to form the prompt
for the LLM as follows:
\begin{enumerate}[start=1,label={(\bfseries \arabic*):},leftmargin=20pt]
    \item \textbf{Identifier to the LLM:} The LLM is told it is a domain
          expert (e.g., a doctor for medical data) and that it is
          provided $n$ reference samples.
    \item \textbf{Reference Samples:} Then, the $n$ reference samples
          are provided in JSON format.
    \item \textbf{Conditioning Values:} Next, the conditioning values
          are provided in human-readable format (e.g., ``age: 45'',
          ``entry time: 10:30'').
    \item \textbf{Generation Instruction:} The LLM is instructed
          to generate a sample in JSON format that is similar to the
          conditioning values and reference samples.
    \item \textbf{Guiding Instructions:} Finally, if necessary, the LLM
          can be given a set of instructions to improve its output
          (e.g., ``If the generated patient would be unrealistic,
          adjust the values slightly'').
\end{enumerate}
The structure of the prompt is designed to minimize context contamination
while aiding the model to generate similar data.
While providing reference samples is not required, not doing so could cause
the synthetic output to not be similar to real data since the model would
miss queues that come from observing the real data.
Then, by placing the reference samples before the conditioning values and
having the instructions last, we ensure that the model focuses on the
conditioning values and remembers the guiding instructions clearly.
A different approach could optimize token caching by placing the guiding
instructions first, and then the context, lowering the computational cost.
We delegate the evaluation of such an approach to future work.

The LLM is then bound by the JSON schema of the dataset to ensure the output
data is parseable through using a grammar-based approach~\cite{llm-grammar-syncode,
    llm-grammar-grammarllm}. This ensures that the output is always valid JSON,
which is especially important for complex schemas and when using local small models.
Finally, the output JSON is parsed and converted to the relational format of
the original data.

\subsection{Realism Evaluation}
A challenge that was faced by this work is the evaluation of the realism
of synthetic data.
As realism is a subjective metric, it is hard to quantify mechanically,
via e.g., statistical metrics.
For this reason, we introduce an unattended evaluation method for realism
based on an LLM agent.

This method re-uses the core aspects of Amalgam's sampling process.
Given a sample, the top $n$ similar samples are collected
from the original dataset using the similarity function defined in
Section~\ref{sec:sampling}.
Then, all $n+1$ samples (the synthetic and the $n$ original) are converted
into JSON.
Next, a prompt is formed for the LLM as follows:
\begin{enumerate}[start=1,label={(\bfseries \arabic*):},leftmargin=20pt]
    \item \textbf{Identifier to the LLM:} The LLM is told it is a
          domain expert and that it is provided $n$
          \emph{real} samples.
    \item \textbf{Real Samples:} The $n$ samples
          are provided as JSON.
    \item \textbf{Task:} The model is told to evaluate the realism of
          an additional sample provided to it and score it from 1 to 5
          for realism.
    \item \textbf{Evaluation Sample:} Finally, the evaluated sample is provided to the LLM.
\end{enumerate}

The LLM is expected to return a score from 1 to 5 and bound by grammar
to ensure the output is parsable.
The score is collected and averaged across multiple samples to provide
an overall realism score for the synthetic dataset.
Then, the evaluation is re-run using a hold-out set of real data samples
that were not used for synthesis or eligible for selection as similar samples.
This provides a realism score for real data, which can be used as a
baseline for the synthetic data.




\section{Experiments}
We compare Amalgam against state-of-the-art in two experiments.
First, we compare Amalgam against MARE~\cite{mare}, a PGM relational
data synthesis algorithm, on four datasets.
Then, we compare Amalgam's performance when using different local LLMs on one
of the datasets.


\subsection{Comparison with State-of-the-Art}
In the first experiment, we compare Amalgam against the state-of-the-art
algorithm MARE~\cite{mare}, which is 
an orchestration algorithm for relational
data using multiple PGMs, according to which each relationship is classified as
independent, associative, or sequential.
We compare to MARE across four datasets: MIMIC-IV Admissions, eICU R1, CTUR SL, and CTUR CE.

MIMIC-IV Admissions is a dataset that consists of three tables in the
core set of MIMIC-IV~\cite{mimic-iv}: ``patients'', ``admissions'', and ``transfers''.
The tables ``admissions'' and ``transfers'' have a sequential relationship,
where ``transfers'' is a child of ``admissions''.
eICU R1 is a dataset that consists of four tables in the eICU dataset~\cite{eicu}:
``patients'', ``admissiondx'', ``vitalaperiodic'', and ``medication'', with an
assortment of columns.
All tables have a child relationship to ``patients''.
The datasets CTUR SL and CTUR CE are sourced from The CTU Prague Relational
Learning Repository~\cite{ctur}, where SL is the StudentLoans dataset,
and CE is the ConsumerExpenditures dataset.
CTUR SL contains key-only tables that show an attribute for a student
(e.g., ``disabled'').
We merge those tables to the main ``person'' table to form a three table
dataset with ``person'', ``enrolled'', and ``enlist'', where ``enrolled''
and ``enlist'' are associative child tables of ``person''.
For CTUR CE, the tables ``household'', ``person'', and
``expenditure'' form a three table dataset.
A dataset overview is shown in Table~\ref{tab:datasets}.
It lists the number of tables, total number of columns, total number
of values (i.e., rows multiplied by columns per table), number of entities
(i.e., rows in the main table), and the average ratio of values to entities.

\begin{table}[h]
    \centering
    \setlength{\tabcolsep}{4pt}
    \begin{tabular}{lrrrrr}
        \hline
        Dataset    & Tbls & Cols & Vals   & N      & Vals/N \\
        \hline
        MIMIC Adm. & 3    & 25   & 16.6M  & 180.7k & 91.6   \\
        eICU R1    & 4    & 25   & 141.3M & 200.9k & 703.5  \\
        CTUR CE    & 3    & 20   & 13.3M  & 56.8k  & 234.5  \\
        CTUR SL    & 3    & 10   & 9.2k   & 1.0k   & 9.2    \\
        \hline
    \end{tabular}
    \caption{Dataset summary statistics.}
    \label{tab:datasets}
\end{table}

In the experiments, we use the same PGM as MARE, PrivBayes~\cite{zhangPrivBayesPrivateData2017b},
with a total privacy budget of $\epsilon=2$.
We retain a 20\% hold-out set for accuracy and realism evaluation.
For accuracy, we collect an aggregated Kullback-Leibler (KL) divergence metric
as shown in MARE~\cite{mare} and a $\chi^2$ metric.
Then, for realism, we use our proposed LLM-based evaluation method.
We use Qwen3 8B~\cite{qwen3} as the LLM for both synthesis and realism evaluation.

For the aggregated KL metric, we collect column pairs in three categories:
intra-table pairs (both columns from the same table), inter-table pairs
(columns from different tables that have a foreign key relationship), and
sequential inter-table pairs (with up to two previous rows).
For each pair in the dataset, we calculate KL divergence between the original
and synthetic (or real) data for each table and normalize them with the formula
$f(x) = 1/(1+x)$, which maps $[0, \infty)$ to $(0, 1]$.
Then, we average the results per category, per-table, and finally we average the
three categories to form the aggregated KL metric.
For the $\chi^2$ metric, we collect the $\chi^2 P$ value per column
between the original and synthetic (or real) data, average them per-table,
and finally average the per-table results to form the final $\chi^2$ metric.
$\chi^2 P$ values indicate the likelihood that two distributions were drawn
from the same underlying distribution, with 100\% being identical distributions.

\begin{figure*}[p]
    \centering
    \begin{subfigure}[b]{0.3\textwidth}
        \includegraphics{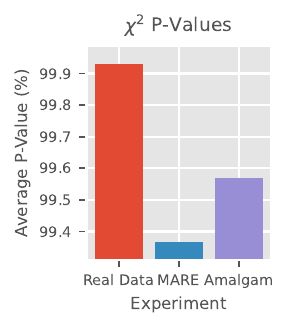}
    \end{subfigure}
    \begin{subfigure}[b]{0.3\textwidth}
        \includegraphics{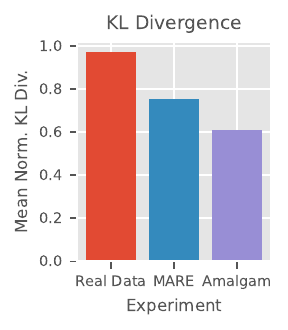}
    \end{subfigure}
    \begin{subfigure}[b]{0.39\textwidth}
        \includegraphics{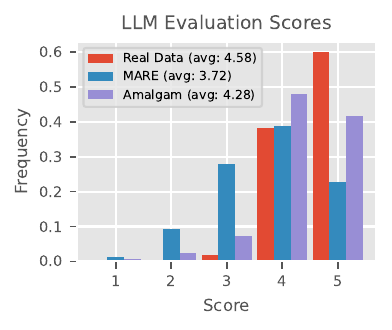}
    \end{subfigure}
    \par\vspace{-1.87em}{\raggedright\small\textbf{MIMIC Adm.}\par}

    \begin{subfigure}[b]{0.3\textwidth}
        \includegraphics{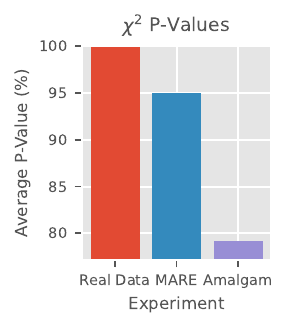}
    \end{subfigure}
    \begin{subfigure}[b]{0.3\textwidth}
        \includegraphics{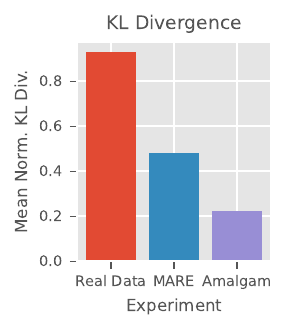}
    \end{subfigure}
    \begin{subfigure}[b]{0.39\textwidth}
        \includegraphics{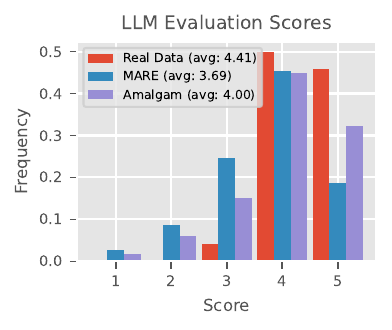}
    \end{subfigure}
    \par\vspace{-1.87em}{\raggedright\small\textbf{eICU R1}\par}

    \begin{subfigure}[b]{0.3\textwidth}
        \includegraphics{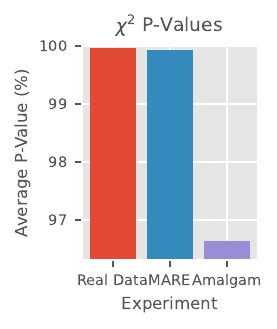}
    \end{subfigure}
    \begin{subfigure}[b]{0.3\textwidth}
        \includegraphics{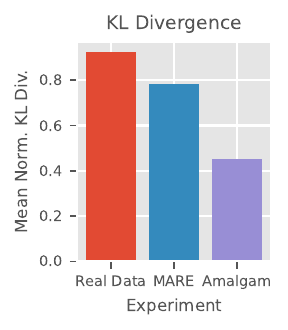}
    \end{subfigure}
    \begin{subfigure}[b]{0.39\textwidth}
        \includegraphics{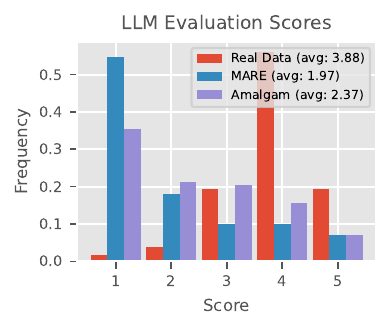}
    \end{subfigure}
    \par\vspace{-1.87em}{\raggedright\small\textbf{CTUR CE}\par}

    \begin{subfigure}[b]{0.3\textwidth}
        \includegraphics{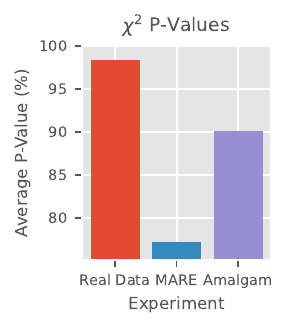}
    \end{subfigure}
    \begin{subfigure}[b]{0.3\textwidth}
        \includegraphics{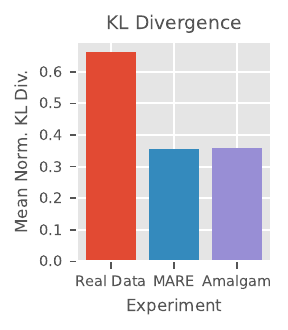}
    \end{subfigure}
    \begin{subfigure}[b]{0.39\textwidth}
        \includegraphics{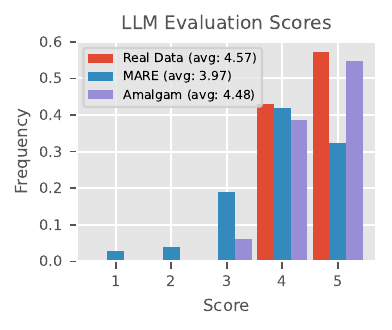}
    \end{subfigure}
    \par\vspace{-1.87em}{\raggedright\small\textbf{CTUR SL}\par}

    \caption{Comparison of Amalgam synthesis results across datasets versus MARE, Real Data.}
    \label{fig:results-combined}
\end{figure*}

Due to the higher computational demand of Amalgam and realism evaluation, we
limit the number of samples that are generated and evaluated for realism to 2000
per dataset (excl.\ CTUR SL where it is 800 total).
MARE always generates the same number of samples as the original dataset,
and accuracy metrics (KL, $\chi^2$) are computed on all produced samples.
The results are shown in Figure~\ref{fig:results-combined}.
Finally, we collect the synthesis time of Amalgam versus MARE for two steps:
structure learning and sampling, shown in Table~\ref{tab:time-comparison}.

\begin{figure*}[tb]
    \centering

    \begin{subfigure}[b]{0.3\textwidth}
        \includegraphics{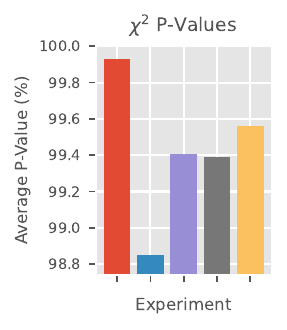}
    \end{subfigure}
    \begin{subfigure}[b]{0.3\textwidth}
        \includegraphics{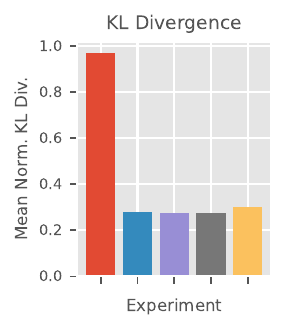}
    \end{subfigure}
    \begin{subfigure}[b]{0.39\textwidth}
        \includegraphics{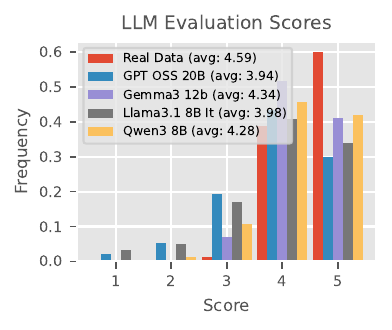}
    \end{subfigure}

    \caption{Comparison of Amalgam synthesis for different LLMs on MIMIC-IV Admissions.}
    \label{fig:results-llm}
\end{figure*}

\begin{figure*}[tb]
    \centering
    \includegraphics[width=0.85\textwidth]{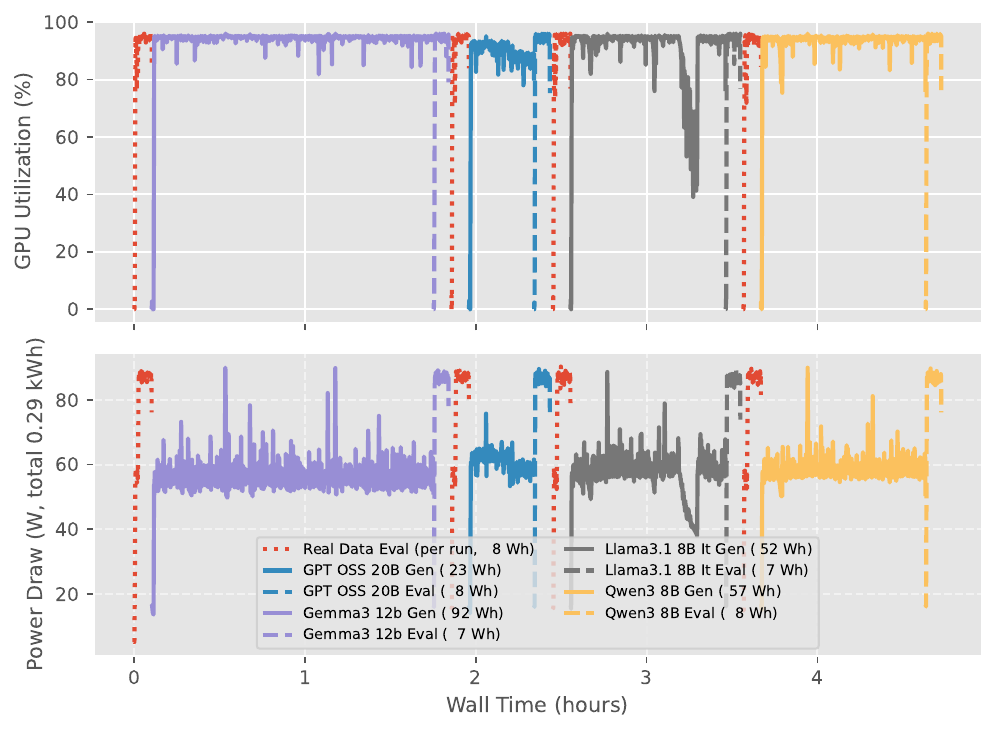}
    \caption{Energy usage comparison across different LLMs on MIMIC-IV Admissions dataset.}
    \label{fig:comparison-energy-llm}
\end{figure*}

\begin{table}[h]
    \centering
    \setlength{\tabcolsep}{4pt}
    \begin{tabular}{lrrrr}
        \hline
                   &
        \multicolumn{2}{c}{MARE}
                   &
        \multicolumn{2}{c}{Amalgam}                 \\
        \cline{2-5}
        Dataset    & Fit   & Sample & Fit  & Sample \\
        \hline
        MIMIC Adm. & 1m29s & 3s     & 1m1s & 8h26m  \\
        eICU R1    & 4m7s  & 3m33s  & 25s  & 12h30m \\
        CTUR CE    & 14.8s & 1s     & 13s  & 6h49m  \\
        CTUR SL    & 11s   & 0.3s   & 4s   & 7m7s   \\
        \hline
    \end{tabular}
    \caption{Time comparison between MARE and Amalgam.}
    \label{tab:time-comparison}
\end{table}

\subsection{LLM Comparison and Efficiency}
In the second experiment, we compare Amalgam using four different LLMs
for synthesis: Qwen3 8B~\cite{qwen3}, Gemma3 12B~\cite{gemma-3},
GPT-OSS 20B~\cite{gpt-oss}, and Meta Llama 3.1 8B Instruct~\cite{meta-llama-3-1}.
Qwen3 8B is used to evaluate realism in all four cases.
We generate 250 samples.
We use the MIMIC-IV Admissions dataset for this experiment and collect
the same metrics as before.
Then, we collect per-second energy and utilization data from the GPU,
running the experiments back to back and recording the relative experiment time (wall time).
The metrics results are shown in Figure~\ref{fig:results-llm}
and the energy analysis in Figure~\ref{fig:comparison-energy-llm}.

\subsection{Hardware Setup}
The experiments are executed on units of Nvidia DGX Spark Founder Edition
that contain a GB10, 128GB of RAM, and 4TB of SSD, using their stock
operating system that mirrors Ubuntu 24.04, with Python 3.12.
PGM models use the CPU and LLM inference the GPU.
For each experiment, we collect per-second energy data from the GPU
using ``nvidia-smi'' and the timing intervals of each synthesis step.

\subsection{Implementation}
We implement Amalgam by extending the data synthesis system Pasteur~\cite{pasteur},
which is written in Python.
Pasteur provides the reference implementation of the algorithm
MARE~\cite{mare} using PrivBayes~\cite{zhangPrivBayesPrivateData2017b},
which is our state-of-the-art reference model.
We implement the structure learning phase by re-using the fitting
process of PrivBayes from Pasteur.
Then, we implement the sampling phase by integrating the LLM
inference framework ``Llama.cpp'' under the library ``llama-cpp-python''
and use the library ``outlines''~\cite{outlines} to perform grammar-based
generation.

We attempted to use online inference providers through OpenRouter
but found that the structured generation via grammar failed when provided
complex schemas for synthesis (e.g., the provider Groq only
performed post-inference validation and discarded more than half of the samples)
and had varying support per provider.
The library ``outlines'' was also pushed to its limits, requiring around 1--3
minutes to prime the grammar for complex schemas and having around a 30\%
GPU throughput degradation compared to normal generation
(shown in Figure~\ref{fig:comparison-energy-llm} power drop).


We source local open-source models from HuggingFace~\cite{huggingface}.
Specifically, we opt to use the model Qwen3 8B~\cite{qwen3} quantized to 4-bit
(``Qwen/Qwen3-8B-GGUF'', filename ``Qwen3-8B-Q4\_K\_M.gguf'') for its
balance of quality and performance.
For comparison, we compare it with models Gemma3 12B~\cite{gemma-3}
(``unsloth/gemma-3-12b-it-GGUF'', filename ``gemma-3-12b-it-Q4\_K\_M.gguf''),
GPT-Oss 20B~\cite{gpt-oss} (``unsloth/gpt-oss-20b-GGUF'', filename
``gpt-oss-20b-Q4\_K\_M.gguf''), and Meta Llama 3.1 8B Instruct~\cite{meta-llama-3-1}
(``MaziyarPanahi/Meta-Llama-3.1-8B-Instruct-GGUF'', filename
``Meta-Llama-3.1-8B-Instruct.Q4\_K\_M.gguf'').
At the time of this writing, only Qwen3 8B provided an official 4-bit GGUF
quantization, so for the other models we used popular community provided
quantizations.

\section{Results}
In the first experiment (Figure~\ref{fig:results-combined}), we observe that
Amalgam provides significantly higher realism than MARE across all datasets,
closing the gap to real data.
Overall, the average realism score across datasets for Amalgam is
3.8/5 compared to MARE's 3.3/5 and real data's 4.4/5.
This is expected, as the LLM is able to leverage its foundational knowledge
to generate more lifelike samples.
Specifically, on MIMIC Admissions and CTUR SL, the Amalgam score distribution
from 1 to 5 overlaps significantly with real data, showing that the synthetic
data is hard to distinguish from real data.
On CTUR CE, the realism is improved significantly over MARE, but there is
a significant number of unrealistic samples for both MARE and Amalgam.
This is likely an effect of the dataset, which uses a lot of arbitrary
product keys under ``expenditures'' that have low semantic meaning.

On accuracy, Amalgam is grounded in reality, with $\chi^2$ scores very close to
100\%, showing high histogram quality.
The exception is eICU R1, where the very large number of rows per entity
causes the LLM to struggle to generate accurate counts for child tables.
The histogram quality was additionally verified manually by observing histogram
plots of the produced columns.
The KL divergence scores are lower, showing that the joint distributions
between columns are not as well preserved.
As the LLM only has access to first row statistics (Section~\ref{sec:fitting})
and not follow-up row statistics, which would be required to preserve joint
distributions, this is expected.
Overall, the $\chi^2$ score across datasets for Amalgam is 91\%,
compared to MARE's 92\%, and for KL divergence, Amalgam scores 0.40
with MARE at 0.60.

Reviewing the times on Table~\ref{tab:time-comparison}, we observe that
Amalgam's structure learning is slightly faster than MARE's, as Amalgam
only fits a single PGM.
However, the sampling time is significantly higher, as sampling uses an LLM.

In the second experiment (Figure~\ref{fig:results-llm}), we observe that
Qwen3 8B and Gemma3 12B provide the best realism scores, with GPT-OSS and
Meta Llama 3.1 8B Instruct lagging behind.
On accuracy, all models provide similar $\chi^2$ scores, with Qwen3 8B
having a slight edge.
Due to the lower number of samples (250), the KL divergence score is negatively
affected compared to the previous experiment,
as rare combinations of values may not appear in the synthetic data.
On the energy axis, we observe that Qwen3 8B uses significantly less energy
than Gemma3 12B while producing slightly better results (Figure~\ref{fig:comparison-energy-llm}),
making it the best choice for synthesis.
GPT-OSS is surprisingly efficient, using two fifths of the energy and runtime
of Qwen3 8B through its Mixture-of-Experts architecture with 3B parameters,
but it performs worse.
MARE runs on CPU only and only requires 1--5m to generate as many samples as
in the original data, so its energy usage is negligible compared to
LLMs.
All models, including real data, are evaluated using Qwen3 8B
for realism, so energy values for evaluation are the same across all models
and real data (hold-out set).

\section{Discussion}
The results show that Amalgam is able to generate synthetic data that
is significantly more realistic than PGM-based synthesis while preserving
high accuracy.
This is achieved by leveraging the foundational knowledge of LLMs
to generate lifelike samples, while using PGMs to ground the generation
in reality.
Through combining these approaches, Amalgam handles complex datasets that
PGM-based synthesis struggles to be realistic with, while avoiding the privacy
and performance issues of LLM fine-tuning.

As Amalgam uses Differential Privacy in the PGM, it ensures that each synthetic
sample is traceable to a fixed number of original samples,
which can be used to audit the synthetic data for privacy violations.
This is a significant advantage over pure LLM-based synthesis with
fine-tuning, where the lack of traceability can lead to privacy concerns.
Methods for automated screening of LLM outputs for privacy violations
is a natural next step for future work.

Moreover, Amalgam runs comfortably on-premises, with all the experiments in
this paper being performed on edge devices (DGX Sparks) with acceptable
runtimes ($\sim7$ hours for 2000 samples), while representing the whole dataset,
which is an excellent attribute for sensitive data synthesis.
Due to its PGM foundation, Amalgam can scale to very large datasets
and fit them in minutes, with its generated data being representative of
the entire input dataset and, if on-demand generation is needed, Amalgam can
begin generating synthetic data a few minutes after cold-start.

However, Amalgam has limitations in certain use-cases where PGMs would be
preferable.
For example, if the goal is purely analytical and the data will not be used
for human consumption, PGM-based synthesis can produce comparable analytics
while being significantly faster and more energy efficient.
In that regard, if the data will be used to stress test systems (e.g., databases),
PGM-based synthesis is a better fit.

In summary, Amalgam has a complementary use-case to PGM-based synthesis,
being preferable when the data will be used for human consumption (e.g.,
application testing) and a large amount of output data is not required.
When analytics are the main goal and large amounts of data are required,
PGM-based synthesis remains preferable.
As an algorithm, Amalgam's resource requirements are relatively low, allowing it
to run on-premises or on edge devices.
In sum, we envision that Amalgam can be used by organizations
in combination with a PGM algorithm, where Amalgam is used to
generate a small set of realistic samples (in the thousands) for UI
stand-ins and PGM synthesis to generate large amounts of accurate samples
(in the hundreds of millions) for e.g., database stress-testing.

\section{Conclusion}
In this paper, we presented Amalgam, a hybrid LLM-PGM synthesis algorithm
that leverages the strengths of LLMs and PGMs to generate lifelike synthetic
data.
Amalgam uses PGMs to learn the joint distribution of key analytics
in the dataset, ensuring that the synthetic data is grounded in reality,
while using pre-trained LLMs to generate samples that are similar to real data.
This combination allows Amalgam to handle complex datasets that PGMs
struggle with, while avoiding the privacy and performance issues of
LLM fine-tuning.
Through experiments on multiple datasets, we show that Amalgam
produces synthetic data that is significantly more realistic than
PGM-based synthesis while preserving high accuracy.
Moreover, Amalgam runs comfortably on-premises, making it suitable
for sensitive data synthesis.


\appendix

\section{Acknowledgments}
This work was supported by the VILLUM Foundation under project ``Teaching AI Green Coding'' (VIL70090); by ITEA4 and the Innovation Fund Denmark for projects ``GreenCode'' (2306) and ``MAST'' (22035).

\section{LLM Prompts}
In this appendix, we provide the prompts used for LLM synthesis for each dataset,
for synthesis and realism evaluation.
In code, the placeholders \texttt{<samples\_n>}, \texttt{<samples>},
\texttt{<seed>}, and \texttt{<eval>} are replaced with the number of samples,
the reference samples in JSON format, the conditioning values in human-readable
format, and the evaluated sample in JSON format respectively.

\subsection{MIMIC-IV Admissions}

\paragraph{Synthesis Prompt}
\begin{quote}
    \small\ttfamily
    You are a doctor working at a hospital.

    Reference the following <samples\_n> example patients:\\
    <samples>

    And derive a JSON patient using:\\
    <seed>

    Guidelines:
    \begin{itemize}[nosep,leftmargin=*]
        \item the provided values are for the first admission/transfer, generate the rest in a realistic manner
        \item for provided ranges use a random value that is within the range
        \item Reason about the key events in this patient's admission; they should make sense
        \item If a value would cause an unrealistic patient, adjust it slightly for realism
    \end{itemize}
\end{quote}

\paragraph{Evaluation Prompt}
\begin{quote}
    \small\ttfamily
    You are a medical doctor.

    You are given the following <samples\_n> real patients as a reference:\\
    <samples>

    Then, you are asked to comment on how real the following patient is with a rating from 1 to 5 (5 being very real):\\
    <eval>
\end{quote}

\subsection{eICU R1}

\paragraph{Synthesis Prompt}
\begin{quote}
    \small\ttfamily
    You are a doctor working at a hospital.

    Reference the following <samples\_n> example patients:\\
    <samples>

    And derive a JSON patient using:\\
    <seed>

    Guidelines:
    \begin{itemize}[nosep,leftmargin=*]
        \item the provided values are for the first admissions etc, generate the rest in a realistic manner
        \item for provided ranges use a random value that is within the range
        \item Reason about the key events in this patient's admission; they should make sense
        \item If a value would cause an unrealistic patient, adjust it slightly for realism
    \end{itemize}
\end{quote}

\paragraph{Evaluation Prompt}
\begin{quote}
    \small\ttfamily
    You are a medical doctor.

    You are given the following <samples\_n> real patients as a reference:\\
    <samples>

    Then, you are asked to comment on how real the following patient is with a rating from 1 to 5 (5 being very real):\\
    <eval>
\end{quote}

\subsection{CTUR CE}

\paragraph{Synthesis Prompt}
\begin{quote}
    \small\ttfamily
    You are a tax accountant.

    Reference the following <samples\_n> households:\\
    <samples>

    And derive a new JSON household using:\\
    <seed>

    Guidelines:
    \begin{itemize}[nosep,leftmargin=*]
        \item the provided values are for the first expense/member, generate the rest in a realistic manner
        \item for provided ranges use a random value that is within the range and 4 decimal places
        \item If a value would cause an unrealistic household, adjust it slightly for realism
    \end{itemize}
\end{quote}

\paragraph{Evaluation Prompt}
\begin{quote}
    \small\ttfamily
    You are a tax accountant.

    You are given the following <samples\_n> real households as a reference:\\
    <samples>

    Then, you are asked to comment on how real the following household is with a rating from 1 to 5 (5 being very real):\\
    <eval>
\end{quote}

\subsection{CTUR SL}

\paragraph{Synthesis Prompt}
\begin{quote}
    \small\ttfamily
    You are a school administrator.

    Reference the following <samples\_n> student profiles:\\
    <samples>

    And derive a new JSON student using:\\
    <seed>

    Guidelines:
    \begin{itemize}[nosep,leftmargin=*]
        \item the provided values are for the first enrollment/enlistment, generate the rest in a realistic manner
        \item for provided ranges use a random value that is within the range and 4 decimal places
        \item If a value would cause an unrealistic student, adjust it slightly for realism
    \end{itemize}
\end{quote}

\paragraph{Evaluation Prompt}
\begin{quote}
    \small\ttfamily
    You are a school administrator.

    You are given the following <samples\_n> real students as a reference:\\
    <samples>

    Then, you are asked to comment on how real the following student is with a rating from 1 to 5 (5 being very real):\\
    <eval>
\end{quote}

\bibliographystyle{icml2026}
\bibliography{llmsynth}

@article{privlava,
  author  = {Kuntai Cai and
             Xiaokui Xiao and
             Graham Cormode},
  title   = {PrivLava: Synthesizing Relational Data with Foreign Keys under Differential
             Privacy},
  journal = {Proc. {ACM} Manag. Data},
  volume  = {1},
  number  = {2},
  pages   = {142:1--142:25},
  year    = {2023}
}

@article{medgan,
  author  = {Kehua Guo and
             Jie Chen and
             Tian Qiu and
             Shaojun Guo and
             Tao Luo and
             Tianyu Chen and
             Sheng Ren},
  title   = {MedGAN: An adaptive {GAN} approach for medical image generation},
  journal = {Comput. Biol. Medicine},
  volume  = {163},
  pages   = {107119},
  year    = {2023}
}

@inproceedings{pategan,
  title     = {PATE-GAN: Generating synthetic data with differential privacy guarantees},
  author    = {Jordon, James and Yoon, Jinsung and Van Der Schaar, Mihaela},
  booktitle = {International conference on learning representations},
  year      = {2018}
}

@inproceedings{privpgm,
  title     = {Graphical-Model Based Estimation and Inference for Differential Privacy},
  booktitle = {Proceedings of the 36th {{International Conference}} on {{Machine Learning}}},
  author    = {Mckenna, Ryan and Sheldon, Daniel and Miklau, Gerome},
  year      = {2019},
  month     = may,
  pages     = {4435--4444},
  publisher = {{PMLR}}
}

@misc{mckennaAIMAdaptiveIterative2022,
  title      = {{{AIM}}: {{An Adaptive}} and {{Iterative Mechanism}} for {{Differentially Private Synthetic Data}}},
  shorttitle = {{{AIM}}},
  author     = {McKenna, Ryan and Mullins, Brett and Sheldon, Daniel and Miklau, Gerome},
  year       = {2022},
  month      = jan
}

@article{caiDataSynthesisDifferentially2021,
  title   = {Data {{Synthesis}} via {{Differentially Private Markov Random Fields}}},
  author  = {Cai, Kuntai and Wei, Jianxin and Lei, Xiaoyu and Xiao, Xiaokui},
  year    = {2021},
  journal = {VLDB}
}

@article{taoBenchmarkingDifferentiallyPrivate2021,
  title  = {Benchmarking {{Differentially Private Synthetic Data Generation Algorithms}}},
  author = {Tao, Yuchao and McKenna, Ryan and Hay, Michael and Machanavajjhala, Ashwin and Miklau, Gerome},
  year   = {2021},
  month  = dec
}

@misc{mckennaWinningNISTContest2021d,
  title  = {Winning the {{NIST Contest}}: {{A}} Scalable and General Approach to Differentially Private Synthetic Data},
  author = {McKenna, Ryan and Miklau, Gerome and Sheldon, Daniel},
  year   = {2021},
  month  = aug
}

@article{zhangPrivBayesPrivateData2017b,
  title   = {{{PrivBayes}}: {{Private Data Release}} via {{Bayesian Networks}}},
  author  = {Zhang, Jun and Cormode, Graham and Procopiuc, Cecilia M. and Srivastava, Divesh and Xiao, Xiaokui},
  year    = {2017},
  month   = oct,
  journal = {ACM Transactions on Database Systems},
  volume  = {42},
  number  = {4},
  pages   = {25:1--25:41}
}

@misc{mimic-iv,
  title  = {{{MIMIC-IV}}},
  author = {Johnson, Alistair and Bulgarelli, Lucas and Pollard, Tom and Horng, Steven and Celi, Leo Anthony and Mark, Roger}
}

@misc{eicu,
  title     = {{eICU Collaborative Research Database}},
  author    = {Pollard, Tom and Johnson, Alistair and Raffa, Jesse and Celi, Leo Anthony and Badawi, Omar and Mark, Roger},
  year      = {2019},
  month     = apr,
  note      = {Version 2.0},
  publisher = {PhysioNet},
  doi       = {10.13026/C2WM1R},
  url       = {https://doi.org/10.13026/C2WM1R}
}

@inproceedings{xuModelingTabularData2019,
  title  = {Modeling {{Tabular}} Data Using {{Conditional GAN}}},
  author = {Xu, Lei and Skoularidou, Maria and {Cuesta-Infante}, Alfredo and Veeramachaneni, Kalyan},
  year   = {2019}
}

@inproceedings{dwork2006differential,
  title  = {Differential Privacy},
  author = {Dwork, Cynthia},
  year   = {2006}
}

@article{llmframework,
  author          = {Goyal, Mandeep and Mahmoud, Qusay H.},
  doi             = {10.1109/CCWC62904.2025.10903878},
  isbn            = {9798331507695},
  journal         = {2025 IEEE 15th Annual Computing and Communication Workshop and Conference, CCWC 2025},
  pages           = {340--346},
  publisher       = {Institute of Electrical and Electronics Engineers Inc.},
  title           = {{An LLM-Based Framework for Synthetic Data Generation}},
  year            = {2025}
}

@article{llmenhance,
  archiveprefix   = {arXiv},
  arxivid         = {2411.03356},
  author          = {Yang, Dayu and Monaikul, Natawut and Ding, Amanda and Tan, Bozhao and Mosaliganti, Kishore and Iyengar, Giri},
  eprint          = {2411.03356},
  month           = {nov},
  title           = {{Enhancing Table Representations with LLM-powered Synthetic Data Generation}},
  url             = {https://arxiv.org/pdf/2411.03356},
  year            = {2024}
}

@article{llmbalance,
  archiveprefix   = {arXiv},
  arxivid         = {2409.19759v3},
  author          = {Chan, Yung-Chieh and Pu, George and Shanker, Apaar and Suresh, Parth and Jenks, Penn and Heyer, John and Denton, Sam},
  eprint          = {2409.19759v3},
  title           = {{Balancing Cost and Effectiveness of Synthetic Data Generation Strategies for LLMs}}
}

@article{transformer,
  author   = {Ashish Vaswani and Google Brain and Noam Shazeer and Niki Parmar and Jakob Uszkoreit and Llion Jones and Aidan N Gomez and Łukasz Kaiser and Illia Polosukhin},
  journal  = {Advances in Neural Information Processing Systems},
  title    = {Attention is All you Need},
  volume   = {30},
  year     = {2017}
}

@article{huggingface,
  title     = {HuggingFace's Transformers: State-of-the-art Natural Language Processing},
  author    = {Wolf, Thomas and Debut, Lysandre and Sanh, Victor and Chaumond, Julien and Delangue, Clement and Moi, Anthony and Cistac, Pierric and Rault, Tim and Louf, R{\'e}mi and Funtowicz, Morgan and Davison, Joe and Shleifer, Sam and von Platen, Patrick and Ma, Clara and Jernite, Yacine and Plu, Julien and Xu, Canwen and Le Scao, Teven and Gugger, Sylvain and Drame, Mariama and Lhoest, Quentin and Rush, Alexander M.},
  journal   = {arXiv preprint arXiv:1910.03771},
  year      = {2020},
  url       = {https://arxiv.org/abs/1910.03771},
  doi       = {10.48550/arXiv.1910.03771}
}

@article{llmquant,
  author   = {Ji Lin and Jiaming Tang and Haotian Tang and Shang Yang and Wei-Ming Chen and Wei-Chen Wang and Guangxuan Xiao and Xingyu Dang and Chuang Gan and Song Han},
  journal  = {Proceedings of Machine Learning and Systems},
  month    = {5},
  pages    = {87-100},
  title    = {AWQ: Activation-aware Weight Quantization for On-Device LLM Compression and Acceleration},
  volume   = {6},
  url      = {https://github.com/mit-han-lab/llm-awq},
  year     = {2024}
}

@article{deepseek-v3,
  author   = {DeepSeek-AI et al.},
  month    = {12},
  title    = {DeepSeek-V3 Technical Report},
  url      = {https://arxiv.org/pdf/2412.19437},
  year     = {2024}
}

@article{gemma-3,
  author   = {GemmaTeam},
  month    = {3},
  title    = {Gemma 3 Technical Report},
  url      = {https://arxiv.org/pdf/2503.19786},
  year     = {2025}
}

@article{llm-grammar-grammarllm,
  author    = {Gabriele Tuccio and Luana Bulla and Maria Madonia and Aldo Gangemi and Misael Mongiovi’},
  doi       = {10.18653/V1/2025.FINDINGS-ACL.177},
  month     = {8},
  pages     = {3412-3422},
  publisher = {Association for Computational Linguistics (ACL)},
  title     = {GRAMMAR-LLM: Grammar-Constrained Natural Language Generation},
  url       = {https://aclanthology.org/2025.findings-acl.177/},
  year      = {2025}
}

@article{llm-grammar-syncode,
  author   = {Shubham Ugare and Tarun Suresh and Hangoo Kang and Sasa Misailovic and Gagandeep Singh},
  issn     = {2835-8856},
  journal  = {Transactions on Machine Learning Research},
  title    = {SynCode: LLM Generation with Grammar Augmentation},
  url      = {https://github.com/uiuc-focal-lab/syncode}
}

@article{mare,
  author   = {Antheas Kapenekakis and Daniele Dell'aglio and Charles Vesteghem and Laurids Poulsen and Martin Bøgsted and Minos Garofalakis and Katja Hose},
  journal  = {IEEE BigData 2024},
  title    = {Synthesizing Accurate Relational Data under Differential Privacy},
  year     = {2024}
}

@article{pgm-koller,
  author    = {Daphne Koller and Nir Friedman},
  editor    = {Thomas Dietterich},
  isbn      = {9780262013192},
  journal   = {Probabilistic Graphical Models: Principles and Techniques},
  pages     = {783-848},
  publisher = {The MIT press},
  title     = {Structure Learning in Bayesian Networks},
  year      = {2009}
}

@inproceedings{dpsgd,
  author    = {Song, Shuang and Chaudhuri, Kamalika and Sarwate, Anand D.},
  booktitle = {2013 IEEE Global Conference on Signal and Information Processing},
  title     = {Stochastic gradient descent with differentially private updates},
  year      = {2013},
  volume    = {},
  number    = {},
}

@InProceedings{pasteur,
  author="Kapenekakis, Antheas
  and Dell'Aglio, Daniele
  and B{\o}gsted, Martin
  and Garofalakis, Minos
  and Hose, Katja",
  editor="Chrysanthis, Panos K.
  and N{\o}rv{\aa}g, Kjetil
  and Stefanidis, Kostas
  and Zhang, Zheying",
  title="Pasteur: Scaling Privacy-Aware Data Synthesis",
  booktitle="Advances in Databases and Information Systems",
  year="2026",
  publisher="Springer Nature Switzerland",
  address="Cham",
  pages="164--180",
  isbn="978-3-032-05281-0"
}

@article{outlines,
  title={Efficient Guided Generation for Large Language Models},
  author={Willard, Brandon T and Louf, R{\'e}mi},
  journal={arXiv preprint arXiv:2307.09702},
  year={2023}
}

@misc{qwen3,
      title={Qwen3 Technical Report},
      author={An Yang and Anfeng Li and Baosong Yang and Beichen Zhang and Binyuan Hui and Bo Zheng and Bowen Yu and Chang Gao and Chengen Huang and Chenxu Lv and Chujie Zheng and Dayiheng Liu and Fan Zhou and Fei Huang and Feng Hu and Hao Ge and Haoran Wei and Huan Lin and Jialong Tang and Jian Yang and Jianhong Tu and Jianwei Zhang and Jianxin Yang and Jiaxi Yang and Jing Zhou and Jingren Zhou and Junyang Lin and Kai Dang and Keqin Bao and Kexin Yang and Le Yu and Lianghao Deng and Mei Li and Mingfeng Xue and Mingze Li and Pei Zhang and Peng Wang and Qin Zhu and Rui Men and Ruize Gao and Shixuan Liu and Shuang Luo and Tianhao Li and Tianyi Tang and Wenbiao Yin and Xingzhang Ren and Xinyu Wang and Xinyu Zhang and Xuancheng Ren and Yang Fan and Yang Su and Yichang Zhang and Yinger Zhang and Yu Wan and Yuqiong Liu and Zekun Wang and Zeyu Cui and Zhenru Zhang and Zhipeng Zhou and Zihan Qiu},
      year={2025},
      eprint={2505.09388},
      archivePrefix={arXiv},
      primaryClass={cs.CL},
      url={https://arxiv.org/abs/2505.09388},
}

@misc{meta-llama-3-1,
  title={The Llama 3 Herd of Models},
  author={Grattafiori, Aaron and Dubey, Abhimanyu and Jauhri, Abhinav and others},
  year={2024},
  eprint={2407.21783},
  archivePrefix={arXiv},
  primaryClass={cs.AI},
  url={https://arxiv.org/abs/2407.21783},
  doi={10.48550/arXiv.2407.21783}
}

@misc{gpt-oss,
  title={gpt-oss-120b \& gpt-oss-20b Model Card},
  author={Agarwal, Sandhini and Ahmad, Lama and Ai, Jason and Altman, Sam and others},
  year={2025},
  eprint={2508.10925},
  archivePrefix={arXiv},
  primaryClass={cs.CL},
  url={https://arxiv.org/abs/2508.10925}
}

@misc{ctur,
    title={The CTU Prague Relational Learning Repository},
    author={Jan Motl and Oliver Schulte},
    year={2024},
    eprint={1511.03086},
    archivePrefix={arXiv},
    primaryClass={cs.LG},
    url={https://arxiv.org/abs/1511.03086},
}

\end{document}